\documentclass[a4paper,12pt,sort&compress]{article}
\usepackage{graphicx}					
\usepackage{subcaption} 
\usepackage[english]{babel}
\usepackage[square,numbers,comma]{natbib}
\usepackage{caption}
\usepackage{amssymb}
\usepackage{amsmath}
\usepackage{amsthm}
\usepackage{bm}
\usepackage{color}
\usepackage{authblk}
\usepackage{mathrsfs}
\providecommand{\keywords}[1]
{
  \small	
  \textbf{\textit{Keywords---}} #1
}
\title{Theorem of resonance of Small Volume High Contrast multilayered materials}
\date{}
\newtheorem{lem:convergence}{Lemma}[section]
\newtheorem{thm:eigenvalueAppr}{Theorem}[section]
\newtheorem{thm:eigenvalueAppr1}{Theorem}[section]
\newtheorem{thm:eigenvalueAppr01}{Theorem}[section]
\newtheorem{thm:eigenvalueAppr02}{Theorem}[section]

\newtheorem{thm:eigenvalueAppr2}{Theorem}[section]
\newtheorem{cor:spcase}{Corollary}[section]
\newtheorem{cor:shari}{Corollary}[section]
\newtheorem{cor:nano}{Corollary}[section]
\author[1]{Taoufik Meklachi, tmeklachi@gmail.com}
\affil[1]{School of Science, Engineering, and Technology, Penn State Harrisburg}
\begin{document}

\maketitle

\begin{abstract}
The need of mathematically formulate relations between composite materials' properties and its resonance response is growing. This is due the fast technological advancement in micro-material manufacturing, present in chips for instance. In this paper two theorems are presented, providing  formulas of scattering resonance of  double-layered and multilayered small volumes in terms of the coefficient of  sussceptibility, being high, and the geometric characteristics. Spectroscopy measurements of the composite medium can exploit the formula to detect its dimension and susceptibility index.              
\end{abstract}

\section{Introduction}
Non-linear spectral analysis of scattering resonances has been a growing field attracting more interest due to the vital applications in imaging, spectroscopy for instance, and material science. In particular, the material's design efficiency at the microscopic level and the refined choice of the  material's components, unarguably,  enhance quality and function of the composite material. In this paper, I will present a first order approximation of scattering resonance of small multilayered volume with high contrast. the layers are concentric and of arbitrary number. Furthermore, the shape of the small volume is also arbitrary, which gives this formula a wide scope of usefulness. The results in this work is a smooth transition from the asymptotic formula for a single small volume with high contrast, elaborated in ref. \cite{Meklachi2018}, to a high contrast small volume composed of multiple concentric media. The scaling technique used in this paper is the same employed in ref. \cite{Meklachi2016}. Data mining and artificial intelligence, from other hand,  were introduced to advance materials manufacturing via an extensive library of materials' properties data. Approach that is especially used when the mathematical physics model is complex to solve or the resulting novel material manifest anomalous phenomena violating physics laws, as in the unnatural bending of incident electromagnetic waves in metamaterials which violates Snell's law ~\cite{Meklachi2016}\cite{Milton2006}\cite{Bouchitt2010}\cite{Bruno2007}\cite{Cai2007}\cite{Greenleaf2009}\cite{Vasquez2009}\cite{Kohn2010}\cite{Kohn2008}\cite{Lai2009}\cite{Liu2009}\cite{McPhedran2009}. For instance, in ref.\cite{Liu2015} machine learning is adopted to study the elastic localization linkages in high-contrast composite materials. A decent number of literature is referenced therein. Most recent contributions to the study of high contrast media can be found in\cite{Meklachi2022}\cite{Ammari2021}\cite{Challa2019}. In section \ref{asymsec} the scattering resonance for two concentric layers is derived, which serves as a basis to generalize the formula to any number of concentric layers in section \ref{asymsec1}.

\section{The double layered case}\label{asymsec}
Consider a small 3D volume $B_h$ in vacuum and contains the origin, with arbitrary shape and a dielectric susceptibility $\eta(x)$ such that $$\eta(x)=\chi_{hB}(x)\frac{\eta_0(x)}{h^2} $$
Suppose $B_h$ is optically inhomogeneous with two concentric high contrast media $B_1$, the inner medium, and $B_2$, the outer one, such that
\[hB=hB_1\cup hB_2, \]
\[\eta(x)=\chi_{hB_1}(x)\eta^1 +\chi_{hB_2}(x){\eta^2}=\chi_{hB_1}(x)\frac{\eta_0^1}{h^2} +\chi_{hB_2}(x)\frac{\eta_0^2}{h^2},\]
and
\[\eta_0(x)=\chi_{B_1}(x)\eta_0^1 +\chi_{B_2}(x){\eta_0^2}\]
for constants $\eta^1$, $\eta^2$, $\eta_0^1$ and $\eta_0^2$.
First, I will derive an asymptotic formula for scattering resonances $\lambda_h$ on the scaled down volume $hB$ with two concentric media. Formula that will be later generalized  to multiple layers of arbitrary number of concentric media. The field $u$ satisfies Helmholtz equation:
\begin{equation}\label{helm}
\Delta u+k^2(1+\eta)u=0
\end{equation}
where $k$ is the wave number and $\lambda=k^2$ is the spectral parameter.
Scaling technique, as performed in \cite{Meklachi2018}, on the integral form of the solution of \eqref{helm} given by Lippmann-Schwinger results in the non-linear eigenvalue problem
\begin{equation}\label{eq:nonlinL}
\lambda_h T_h(\lambda_h)u_h=u_h
\end{equation}  
where
\begin{align}\label{eq:Th1}
    T_h(\lambda)(u)({x})&=\frac{\eta_0^1}{4\pi}\int_{B_1}\frac{\exp({i\sqrt{\lambda}h|x-y|)}}{|x-y|}u(y)dy\\ &+\frac{\eta_0^2}{4\pi}\int_{B_2}\frac{\exp({i\sqrt{\lambda}h|x-y|)}}{|x-y|}u(y)dy
\end{align}

The limiting form of equation \eqref{eq:nonlinL}
\begin{equation}\label{eig1} 
\lambda_0 T_0(u_0)=u_0
\end{equation}
as $h \rightarrow 0$ is a linear eigenvalue problem where 
\begin{equation}\label{eq:T00}
    T_0(u)({x})=\frac{\eta_0^1}{4\pi}\int_{B_1}\frac{1}{|x-y|}u(y)dy+\frac{\eta_0^2}{4\pi}\int_{B_2}\frac{1}{|x-y|}u(y)dy
\end{equation}
Let \[U_1=\int_{B_1}u_0(x)dx \quad \text{and    }\quad U_2=\int_{B_2}u_0(x)dx\] 
The following Theorem provides a first order approximation to $\lambda_h$ for small volume high contrast two concentric media.
\begin{thm:eigenvalueAppr01}\label{thm5}
Let $U$ be a domain bounded  away from the negative real axis in $\mathbb{C}$. Let $T_0$ and $T_h(\lambda)$  be two linear compact  operators  from $L^2(B)$ to $L^2(B)$ defined by \eqref{eq:T00} and \eqref{eq:Th1}, respectively. Let $\lambda_0\neq0$ in $U$ be a simple eigenvalue of $T_0$, and let $u_0$ be the normalized eigenfunction. Then for $h$ small enough, there exists a nonlinear eigenvalue $\lambda_h$ of $T_h$ satisfying the formula:
\begin{equation}\label{asymptotic2lay}
	\lambda_h=\lambda_0-i\frac{{\lambda_0}^\frac{5}{2}}{4\pi}\left[{\eta_0^1U_1^2+\eta_0^2U_2^2+(\eta_0^1+\eta_0^2)U_1U_2}\right]h+\mathcal{O}(h^2).
\end{equation}  
\end{thm:eigenvalueAppr01}
\begin{proof} This proof uses Lemma 2.1 and Theorem 2.1 in ref. \cite{Meklachi2018}.
First, $T_h(\lambda)$ converges uniformly in the $L^2(B)$ norm to $T_0$ as $h\rightarrow0$. This can be shown by using Lemma 2.1 to first establish  uniform converge on $L^2(B_1)$ and $L^2(B_2)$, then the triangular inequality property of the $L^2$-norm establishes the uniform convergence on $L^2(B)$. In fact, there exist positive $C_1$ and $C_2$ such that
\begin{align*}
    ||T_h(\lambda)-T_0||&\leq||(T_h(\lambda)-T_0)|_{B_1}+(T_h(\lambda)-T_0)|_{B_2}||\\
    &\leq||T_h(\lambda)-T_0||_{L^2(B_1)}+||T_h(\lambda)-T_0||_{L^2(B_2)}\\
    &\leq C_1h+C_2h \quad \text{by Lemma 2.1}\\
    &=(C_1+C_2)h
\end{align*}

Theorem 2.1 in \cite{Meklachi2018} provides a first order approximation of $\lambda_h$ given by 
\begin{equation}\label{th2.1form}
	\lambda_h=\lambda_0+{\lambda_0}^2\large\left\langle(T_0-T_h(\lambda_0))u_0,u_0\large\right\rangle+\mathcal{O}(h^2)
\end{equation} 
In particular we have 
\begin{align}\label{display}
	(T_0-T_h(\lambda))(u_0)(x)&=\frac{\eta_0^1}{4\pi}\int_{B_1}\big(1-\exp({i\sqrt{\lambda}h|x-y|)}\big)  \frac{u_0(y)}{|x-y|}dy\\
	&+\frac{\eta_0^2}{4\pi}\int_{B_2}\big(1-\exp({i\sqrt{\lambda}h|x-y|)}\big)  \frac{u_0(y)}{|x-y|}dy\label{display2}
\end{align}   
Taylor expansion on the function  $h\rightarrow \exp({i\sqrt{\lambda}h|x-y|)}$ to the first order gives
\[\big(1-\exp({i\sqrt{\lambda}h|x-y|)}\big)  \frac{1}{|x-y|}=-i\sqrt{\lambda}h+\mathcal{O}(h^2)\]
Substituting in \eqref{display} and \eqref{display2} we obtain
\begin{equation*}
 (T_0-T_h(\lambda))(u_0)(x)=-i\frac{{\lambda_0}^\frac{1}{2}}{4\pi}(\eta_0^1U_1+\eta_0^2U_2)h+\mathcal{O}(h^2)
\end{equation*}
 Finally plugging the last expression in \eqref{th2.1form} concludes formula \eqref{asymptotic2lay} 
\end{proof}
A useful formulation in spectroscopy applications when computing the volume of $B_h$ could be
\begin{equation}\label{asymptotic2lay1}
	\lambda_h=\lambda_0-i\frac{{\lambda_0}^\frac{5}{2}}{4\pi}\left[{(\eta_0^1U_1+\eta_0^2U_2)U_0}\right]h+\mathcal{O}(h^2).
\end{equation}  
where \[U_0=\int_{B}u_0(x)dx.\]
The special case when $\eta_0^1=\eta_1^2$, hence $\eta^1=\eta^2$, provides exactly the asymptotic formula for the resonance derived for a single small volume and high contrast scatterer, ref. \cite{Meklachi2018}, which writes
\begin{equation}\label{asymptotic}
	\lambda_h=\lambda_0-i\frac{\eta_0}{4\pi}{\lambda_0}^\frac{5}{2}{U_0}^2h+\mathcal{O}(h^2).
\end{equation}
\section{The multilayered resonance Theorem}\label{asymsec1}
Let $B_h$ be a small 3D volume  that contains the origin with arbitrary shape and dielectric susceptibility $\eta(x)$ such that $$\eta(x)=\chi_{hB}(x)\frac{\eta_0(x)}{h^2} $$
Suppose $B_h$ is composite and optically inhomogenous with $n$ layers of concentric high contrast media ${\{B_i}\}_{1\leq i\leq n}$, such that
\[hB=\cup_{i=1}^{i=n} hB_i, \]
\[\eta(x)=\sum_{i=1}^{i=n}\chi_{hB_i}(x)\eta^i =\sum_{i=1}^{i=n}\chi_{hB_i}(x)\frac{\eta_0^i}{h^2},\]
and
\[\eta_0(x)=\sum_{i=1}^{i=n}\chi_{B_i}(x)\eta_0^i\]
for constants $\eta^i$, $\eta_0^i$. The field $u$ satisfies Helmholtz equation \eqref{helm}, and similarly to the double layer case, can be formulated as an integral non-linear eigenvalue problem \eqref{eq:nonlinL} where 
\begin{equation}
     T_h(\lambda)(u)({x})=\sum_{k=1}^{k=n}\frac{\eta_0^k}{4\pi}\int_{B_k}\frac{\exp({i\sqrt{\lambda}h|x-y|)}}{|x-y|}u(y)dy
\end{equation}
having the limiting linear eigenvalue problem \eqref{eig1} of the operator 
\begin{equation}
    T_0(u)({x})=\sum_{k=1}^{k=n}\frac{\eta_0^k}{4\pi}\int_{B_k}\frac{1}{|x-y|}u(y)dy
\end{equation}
as $h\rightarrow 0$.
\begin{thm:eigenvalueAppr02}
Assume that the hypotheses in Theorem \eqref{thm5} hold. The scattering resonance satisfying \eqref{eq:nonlinL} is given by
\begin{equation}\label{asymptotic2lay2}
	\lambda_h=\lambda_0-i\frac{{\lambda_0}^\frac{5}{2}}{4\pi}\left[{U_0\sum_{k=1}^{k=n}\eta_0^kU_k}\right]h+\mathcal{O}(h^2)
\end{equation}  
where  \[U_0=\int_{B}u_0(x)dx,\]
and
$$U_k=\int_{B_k}u_0(x)dx, \quad 1\leq k\leq n.$$
\end{thm:eigenvalueAppr02}
\begin{proof}
Uniform converge of $T_h(\lambda)$ to $T_0$ as $h \rightarrow 0$ can be shown the same way as in the proof of Theorem \eqref{thm5}. Furthermore, and similarly to the two layer case, we have
\begin{align*}
    (T_0-T_h(\lambda))(u_0)(x)&=\sum_{k=1}^{k=n}\frac{\eta_0^k}{4\pi}\int_{B_k}\big(1-\exp({i\sqrt{\lambda}h|x-y|)}\big)  \frac{u_0(y)}{|x-y|}dy\\
    &=-i\frac{{\lambda_0}^\frac{1}{2}}{4\pi}(\sum_{k=1}^{k=n}\eta_0^kU_k)h+\mathcal{O}(h^2)
\end{align*}
which, substituted in formula \eqref{th2.1form}, gives the asymptotic expression \eqref{asymptotic2lay2} 
\end{proof}
\newpage

\bibliographystyle{siam.bst}
\bibliography{Spectral2.bib}

\end{document}